\documentclass[preprint2]{aastex}

\slugcomment{Not to appear in Nonlearned J., 45.}

\shorttitle{Synthetic Observations of Diffuse Multiphase Clouds}
\shortauthors{Yamada et al.}

\begin{document}

\title{Synthetic Observations of Carbon Lines of Turbulent Flows \\
       in Diffuse Multiphase Interstellar Medium}

\author{M. Yamada\altaffilmark{1}}
\affil{Division of Theoretical Astronomy, National Astronomical
  Observatory of Japan, Mitaka, Osawa, 181-8588, JAPAN}
\email{ymasako@th.nao.ac.jp}

\author{H. Koyama\altaffilmark{2}}
\affil{Department of Astronomy, University of Maryland,
        College Park, MD 20742}

\author{K. Omukai\altaffilmark{1}}

\and

\author{S. Inutsuka\altaffilmark{3}}
\affil{Department of Physics, Kyoto University, Kyoto 606-8502, JAPAN}

\begin{abstract}
We examine observational characteristics of multi-phase turbulent
flows in the diffuse interstellar medium (ISM) using a synthetic
radiation field of atomic and molecular lines.
We consider the multi-phase ISM which is formed by thermal instability
under the irradiation of UV photons with moderate visual extinction
$A_V\sim 1$.
Radiation field maps of C$^{+}$, C$^0$, and CO line emissions were
generated by calculating the non-local thermodynamic equilibrium
(nonLTE) level populations 
from the results of high resolution hydrodynamic simulations of
diffuse ISM models.
By analyzing synthetic radiation field of carbon lines of [\ion{C}{2}]
158 $\mu$m, [\ion{C}{1}] $^3P_2-^3P_1$ (809 GHz), $^3P_1-^3P_0$ (492
GHz), and CO rotational transitions, we found a high ratio between
the lines of high- and low-excitation energies in the diffuse
multi-phase interstellar medium. 
This shows that simultaneous observations of the lines of 
warm- and cold-gas tracers will be useful in examining the thermal
structure, and hence the origin of diffuse interstellar clouds.
\end{abstract}


\section{Introduction}
Recent studies of the multi-phase interstellar medium (ISM) have
 demonstrated that the origin of small scale interstellar turbulence 
 might be attributed to the thermally unstable nature of the ISM
 itself \citep{koyama02,
 koyama06,kritsuk02a,kritsuk02b,audit05,heitsch05,inutsuka05,
 VazquezSemadeni06}. 
Until recently, most theoretical work on interstellar turbulence has
 assumed that the ISM is isothermal (see for recent reviews,
 \citealt{maclow04};  \citealt{elmegreen04}, and references therein).
This approximation has been imposed mainly because quiescent gas
 evolves almost isothermally at 10K as a result of the thermal
 balance between radiative heating and dust emission cooling where
 external UV radiation is well-shielded
 \citep*{hattori69}, and partly because of limited computational
 ability.
Detailed analyses of the one-dimensional collapse due to thermal 
 instability can be found in, for example, 
 \citet{hennebelle99, hennebelle00} and \citet{koyama00}.  
The situation has changed since \citet{koyama02} first pointed out 
 that the thermally unstable nature of multi-phase ISM is able 
 to maintain the turbulent flows. 
Since then, various attempts have been made to examine the properties
 of the multi-phase turbulent ISM \citep{koyama02, kritsuk02a,
 kritsuk02b, audit05}. 
\citet{koyama02} performed two-dimensional simulation of a
 supernova-induced shock propagating into warm ($\sim$ 6000K) ISM.
They demonstrated that broad emission linewidths can be generated
 in the post-shock layer, where chaotic structures are formed by
 thermal instability.
In their calculation, tiny, cold and dense clumps were seen to move in
 the surrounding warm diffuse medium subsonically with respect to the
 warm medium, but supersonically to the sound speed in the cold clumps.
In this article, we refer to this model as the ``two-phase model'' 
 where two thermal states, i.e., cold clumps and warm medium, 
 are generated by thermal instability.
Similar conclusions have also been reached by other authors.
Among them, \citet{kritsuk02a, kritsuk02b} performed three dimensional
 simulations of a multi-phase medium and examined the evolution of
 kinetic energy of the resulting turbulence.
\citet{audit05} simulated a two-dimensional converging flow, where the
 post-shock thermally unstable gas is continuously provided by the two
 shock waves that confine the post-shock gas. 
They showed that complex structures similar to those of 
 \citet{koyama02} are formed, and discussed their evolution by 
 comparing cooling length, 
 fragment scale of the cold neutral medium (CNM), Field length, 
 and the typical size of the shocked layer.
\citet{heitsch05} and \citet{VazquezSemadeni06} have 
 followed this line of work and discussed the observational 
 signature as well. 
While the calculations above focus mainly on neutral atomic gas,
 \citet{pavlovski05} examined molecular gas of much higher density
 ($n_{\rm H} \simeq 10^6$ cm$^{-3}$).
They concluded that the gas is kept isothermal due to the fast
 reformation of the coolant molecules behind shock waves 
 in the dense molecular clouds.

In this article, we concentrate on more diffuse gas 
 ($\approx 1$ cm$^{-3}$), which is warmed by mild UV photoelectric 
 heating, and focus on observational 
 characteristics of the two-phase medium.
Emission-line diagnosis of atomic and molecular radiation in 
 millimeter and submillimeter bands provides a powerful tool 
 for observational study of interstellar turbulence 
 \citep[e.g.,][]{padoan01,bergin04}.
In addition to existing telescopes equipped with submillimeter
 receivers, such as SubMillimeter Array (SMA) and Atacama Submillimeter
 TElescope (ASTE), the highly sensitive instruments of Atacama Large
 Millimeter/submillimeter Array (ALMA) will be available before long.
We thus examine how a characteristic feature of the two-phase 
 turbulence model should appear in observational quantities.
We try to find and propose unambiguous and robust
 observational indicators for the two-phase turbulence model.
For this purpose, we perform hydrodynamical simulations of 
 two-phase turbulent flows, and then calculate line emission of atomic 
 carbons and CO molecules.
Based on the synthetic emission maps, 
 our results are more suitable for direct comparison with observation 
 than the density or temperature distributions, 
 which are provided by previous hydrodynamical simulations.

The organization of this article is as follows:
In \S 2 we describe our model and basic equations.
In \S 3 we display our synthetic radiation maps of the two-phase model
as well as the conventional one-phase (isothermal) turbulent flow model.
We also show the results of the analyses of our simulated emission maps
 of C$^{+}$, C$^0$, and CO lines.
In \S 4 we discuss the limits of and uncertainty about our calculation 
 and discuss its relevance to current observational results.
We summarize this article in the final section (\S 5).

\section{Models and Calculations}

We consider diffuse clouds irradiated by mild UV background radiation
of $G_0=1$ (where $G_0$ is the typical interstellar UV strength in
Habing unit), which is attenuated with average visual extinction
$A_V\lesssim$ 0.1. 
We perform hydrodynamical simulations first, and then calculate 
line-emissivity distributions using the simulation data. 
In the following, we present equations and assumptions for hydrodynamical 
and radiation calculations in this order.

\subsection{Hydrodynamic Equations}

We solve an ordinary set of hydrodynamical equations:
  \begin{mathletters}
  \begin{eqnarray}
    \frac{\partial\rho}{\partial t} 
       +\nabla\cdot (\rho \boldmath{v}) &=& 0, \\
    \frac{\partial\boldmath{v}}{\partial t}
       +(\boldmath{v}\cdot\nabla)\boldmath{v} 
       &=& -\frac{\nabla P}{\rho}, 
                                                \label{eq:EoM} \\
    \frac{\partial\rho e}{\partial t}
       +\nabla\cdot (\rho e\boldmath{v}) +P\nabla\cdot\boldmath{v}
       &=& \frac{\rho}{m_H}\Gamma
           -\left( \frac{\rho}{m_H}\right)^2\Lambda(T) \nonumber \\
	& &   +\nabla\cdot(K\nabla T) 
                                                \label{eq:energy}
  \end{eqnarray}
  \end{mathletters}
where $\rho$ is the density, $P$ is the pressure, $e=P/(\gamma-1)$ is the 
specific internal energy with the adiabatic index $\gamma=5/3$,
$\Gamma$ and $\Lambda$ represent the heating and cooling rates,
respectively, and $\boldmath{v}$ is the velocity of a fluid element. 
The last term of the third equation (\ref{eq:energy}) represents 
thermal conduction. 
The thermal conduction coefficient 
$K=2.5\times 10^3T^{1/2}$ ergs cm$^{-1}$ K$^{-1}$ sec$^{-1}$ is taken
from \citet{parker53}, which is valid for temperature $T\lesssim
4.47\times 10^4$K.
For cooling and heating rates, we employ the fitting formulae 
 shown by equation (8) of \citet{koyama06}. 
The formulae are derived by approximately describing the result of
\citet{koyama00}, who performed non-equilibrium thermal calculations
of one-dimensional shock waves.
Their formulae contain dust-grain photoelectric heating, 
 cosmic ray and X-ray ionization heating, H$_2$ formation and
 destruction, and line cooling by atomic and molecular emission.
In equation (\ref{eq:EoM}) we do not introduce the physical viscosity,
which affects the saturation level of turbulent motions, because we
are more interested in characteristic features in radiation field
rather than detailed arguments on the mechanism for maintaining the
turbulent motion.
Detailed systematic analyses of the saturation amplitude of two-phase
turbulent motions are found in \citet{koyama06}.
We can safely ignore back reactions by cooling radiation to
hydrodynamical evolutions either via radiation pressure or radiation
force.

Our two-dimensional simulations employ a scheme based on 
 the second-order Godunov method. 
In order to correctly resolve thermal conduction-induced structures, 
the Field length $\lambda_\mathrm{F}=(KT/n^2\Lambda)^{1/2}$, 
which is the characteristic length scale of thermal conduction, 
must be resolved by more than three cells (the so-called Field
 condition, after \citealp{koyama04}).
Our calculations satisfy this condition with the number of Eulerian 
grids 2048$^2$ in the simulation box of 2.4$\times$2.4 pc$^2$.

We introduce a supersonic velocity field to the medium of initially
uniform density and temperature, and calculate its evolution without
an external driving force (i.e., so-called decaying turbulence). 
The initial supersonic velocity field has an average Mach number of 
 $\langle \mathcal{M}\rangle\approx 10$, and power spectrum
 $P(k)\propto k^{-6}$ in the range of 
 $1 \le k/2\pi=1/\lambda \le 4$, in units of
 the reciprocal of the size of the simulation box $\lambda$.
The purpose of this paper is to simulate the radio band observation 
 of certain samples of turbulent clouds. 
Thus we set up the initial conditions rather arbitrarily and 
 calculate the dynamical evolution of the relaxation process 
 from these highly turbulent flows.  
For the actual simulation of the observation described below, 
 we choose snapshots of the time evolution that would mimic
 the observed velocity profiles (i.e., position-velocity diagram) 
 of diffuse clouds.

\subsection{Method of Radiation Calculation}

We solve non-local thermodynamic equilibrium (nonLTE) level
populations for relevant lines of atomic and molecular carbon,
C$^{+}$, C$^{0}$, and CO.
Each level population is determined by solving the equations of
detailed balance for transitions as a function of local density and
temperature taken from the hydrodynamical calculations.
In our model of mildly UV-irradiated diffuse medium, in addition to
the Ly$\alpha$ line of atomic hydrogen in the UV band, various
millimeter and submillimeter lines contribute to the cooling. 
Among the line-cooling processes, the contribution of [\ion{C}{2}]
fine structure cooling is dominant in a wide density range of diffuse
regimes \citep[see Figure 1a of ][]{koyama00}. 
Toward higher density, the role of C$^{0}$ and CO cooling gradually
increases. 
Around the column density $N_{\rm H}=10^{20}$cm$^{-2}$ where
background UV radiation of $G_0=1$ begins to be shielded, the dominant
coolant changes from ionic to molecular carbon. 
Our ISM model, which includes heating and cooling under
$N_H=10^{20}$cm$^{-2}$, corresponds to the atomic/molecular transition
region, and thus ionic, neutral and molecular carbon species should
exist simultaneously. 
Therefore we examine three species of carbon, i.e., C$^{+}$, C$^{0}$,
and CO, in order to probe various thermal states in
multi-phase ISM. 
We calculate \ion{C}{2} and \ion{C}{1} fine structure
transitions, whose excitation energies are 92K([\ion{C}{2}]),
24K([\ion{C}{1}]$^3P_1-^3P_0$), 39K([\ion{C}{1}]$^3P_2-^3P_1$), and
63K([\ion{C}{1}]$^3P_2-^3P_0$), as well as CO($J-$($J-1$)) rotational
transitions, with $5.5 J$ K, where $J$ is the rotational quantum
number.

For [\ion{C}{2}] 158$\mu$m line, we use a modified cooling function by
\citet{koyama02};
  \begin{eqnarray}
    \Lambda_\mathrm{CII} &=&
     2.8\times 10^{-28}\sqrt{T}\cdot\exp\left( -\frac{92}{T}\right) \nonumber \\
    & & \times y(\mathrm{C}^{+})[y({\mathrm H}^0)+y({\mathrm{H}_2})] 
        {n_{\rm H}}^2  \nonumber \\
    & & ~ ~ ~ ~ ~ ~ \mathrm{erg}~ \mathrm{sec}^{-1}\mathrm{cm}^{-3}, 
              \label{eq:fitting}
  \end{eqnarray}
where $y(X)=n(X)/n_{\rm H}$ represents the fractional abundance of the 
line-emitting particle $X$.
Both atomic and molecular hydrogen work as collision partners for
exciting C$^{+}$. 
In the regime where \ion{C}{2} cooling dominates, we can safely
replace the term for collision partners with $y({\mathrm
  H}^0)+y({\mathrm{H}_2})=y({\mathrm H}^0)$.

The emissivities per unit volume by [\ion{C}{1}] and CO lines are
calculated by solving the nonLTE level population.
The emissivity of species $X$ by transition from level $i$ to $j$ per
unit volume (cooling rate) is given by 

  \begin{equation}
    \Lambda_{X,ij}= h\nu_{ij} A_{ij} n(X,i), \label{eq:coolingrate}
  \end{equation} 
where $h\nu_{ij}$ is the energy difference between levels $i$ and $j$, 
$A_{ij}$ is the Einstein A-coefficient for spontaneous emission from
energy level $i$ to $j$, and $n(X,i)$ is the population density of
species $X$ and energy level $i$.
The level populations are obtained by solving the equations of
statistical balance:
  \begin{equation}
    n(X,i) \sum_{j'} (A_{ij'}+ C_{ij'})= \sum_{j'} n(X,j')[A_{j'i}+C_{j'i}].
  \end{equation}
In the above equation, $C_{ij}$ is the collisional transition rate per 
emitting particle, which is written with the collisional rate
coefficient $\gamma_{ij}$ as 
  \begin{equation}
    C_{ij}=\sum_{Y} n(Y) \gamma_{ij},
  \label{eq:Cij}
  \end{equation}
where $n(Y)$ indicates the density of collision partner $Y$.
For neutral carbon atoms, we consider the triplet 
in the electronic ground state ($^3P_0,^3P_1,^3P_2$). 
Relevant transition coefficients are taken from \citet{hollenbach89}. 
The collision partners for excitation of atomic carbon include atomic
hydrogen and electron, and neglect H$_2$ for only a small contribution
of H$_2$ is expected where \ion{C}{1} abundance and line emission have
their peaks (the abundance of C$^0$ reaches its maximum between the
surface of UV irradiated region and the well-shielded inner region
where most of carbon is locked in CO molecules). 
For CO molecules, we make use of a simplified formula for the
collisional rate coefficients of \citet{mckee82}.
CO is assumed to be excited by collision with atomic and
molecular hydrogen.
We use the same collisional coefficient as atomic hydrogen for H$_2$.

In order to calculate line intensities we need to know the ionization
degree and the chemical abundances for each species. 
This requires non-equilibrium calculation of chemical evolution  
 along with hydrodynamical simulation, 
 which is computationally very expensive. 
To avoid this difficulty, here we simply assume uniform abundance
 distribution, and a prescribed variation in terms of temperature (see
 \S \ref{sect:COabund}). 
For the ionization degree, we assume an electron fraction of
 $y(e)=10^{-2}$ referring to the thermal equilibrium
 abundances of \citet{wolfire95}. 
In the neutral medium we are considering, change in the assumed value 
 of $y(e)$ does not significantly alter the results 
 since the collisional transition by electrons is not dominant.
For carbon species, we take $y({\mathrm{C}^{+}}) =
 10^{-3}$ and $y({\mathrm{C}^{0}}) =y(\mathrm{CO}) =3\times 10^{-7}$. 
These abundances are taken from typical values in the one-dimensional 
 calculation of \citet{koyama00} except for ${\mathrm{C}^{0}}$. 
The abundance of ${\mathrm{C}^{0}}$ was not calculated in
 \citet{koyama00}, and we arbitrarily assumed the same value as CO for
 simplicity.
The fractions of C$^{0}$ and CO are affected by such factors as 
 the degree of shielding of ultraviolet radiation and the abundances
 of other species.
We try to derive quantities that are independent of the abundance
 distribution, and defer discussion on the effects of chemical
 evolution in our model to later in \S\ref{sect:COabund}.

Line emissivity of each fluid element has the thermally broadened
profile
  \begin{equation}
    \phi(\nu) = \frac{1}{\Delta\nu_D\sqrt{\pi}}
      \exp \left( -\frac{(\nu-\nu_0)^2}{(\Delta\nu_D)^2}\right), 
      \label{eq:profile}
  \end{equation}
where $\nu_0$ denotes the frequency of the line center, and
$\Delta\nu_D=(\nu_0/c) \sqrt{2k_BT/m}$ denotes the thermal width at
 temperature $T$. 
Line profiles are obtained by convolving the thermal
 broadening with the line centroid $\nu_0$ Doppler-shifted by the bulk
 fluid motion.
The bulk fluid velocity is taken from results of the hydrodynamical
 calculations.

\section{Results}

\subsection{Line Emission Maps}

\begin{figure}[thbp]
\epsscale{1.0}
\plotone{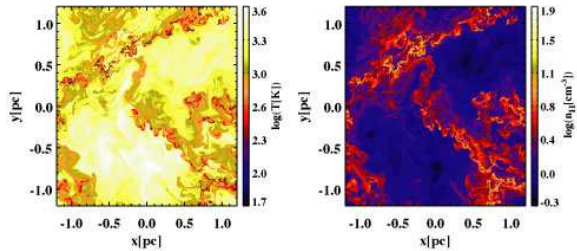}
\caption{(a) Temperature and (b) density distributions of the
  two-phase flow model at time $t=2.479$ Myr.
  Numerous tiny dense and cold clumps are formed within the simulation
  box.
  \label{fig:map_0012}}
\end{figure}

\begin{figure}[htbp]
\epsscale{1.0}
\plotone{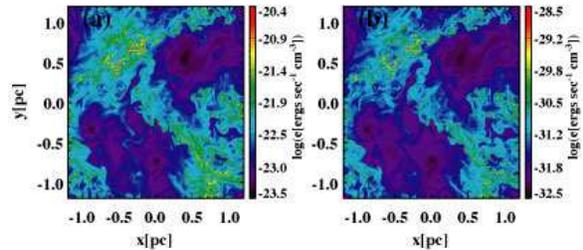}
\caption{Synthesized (a)[\ion{C}{2}] and (b) CO$(1-0)$ line-emissivity 
  distributions in the two-phase flow model at $t=$ 2.479 Myr.
  Intensities of the emission lines, which are in proportion to
  the emissivity in the optically thin case, reflect both 
  the density and temperature distributions.
  Radiation is emitted mainly from tiny, cold, dense clumps.
  \label{fig:linemapF_0012}}
\end{figure}

Figure \ref{fig:map_0012} shows a snapshot of 
 temperature and density distributions. 
In the density distribution map (Figure \ref{fig:map_0012}b), numerous
tiny dense clumps are observed, having been formed as a consequence of
phase transition in the thermally unstable medium after the initial
 shock heating. 
Comparing Figures \ref{fig:map_0012}a and b, we can easily recognize
 that the density map is just a color-inverted version of the
 temperature map. 
This is due to the fact that this turbulent flow is essentially 
 driven by the thermal instability characterized by its approximate
 isobaricity.
Under almost uniform pressure, thermally unstable gas forms numerous
 tiny cold and dense clumps within the surrounding warm and diffuse gas
 \citep{koyama02,kritsuk02a,audit05}. 
Figures \ref{fig:linemapF_0012}a and \ref{fig:linemapF_0012}b
represent the synthesized emissivity maps of [\ion{C}{2}] 158 $\mu$m
 and CO$(1-0)$ lines.
Since the density is below the critical density for LTE $n_\mathrm{crit}
(\approx 3000 $cm$^{-3})$ throughout the whole region, and both lines
 are assumed to be optically thin, line emissivity per unit volume is
proportional to $n_{\rm H}^2$ under the assumption of uniform
 abundance distribution.
Thus the emissivity map closely resembles the density distribution one.
Tiny cold clumps strongly emit radiation in both [\ion{C}{2}] and
CO$(1-0)$ lines because of high density.

\begin{figure}[bhtp]
\epsscale{.60}
\plotone{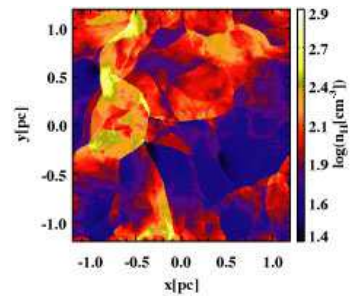}
\caption{Density distribution in the one-phase flow model at $t=5.248$
  Myr. Woven structures with sharp edges are observed, formed through
  strong shock compression by initial supersonic turbulent motions.
  \label{fig:map_0004}}
\end{figure}

\begin{figure}[htbp]
\epsscale{1.0}
\plotone{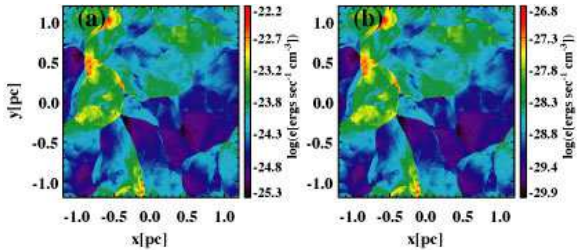}
\caption{Emissivity distributions of (a) [\ion{C}{2}] 
  and (b) CO$(1-0)$ lines in the one-phase flow model with 
  $T=10$K at $t=5.248$ Myr. 
  Highly woven structures are also seen in line-intensity
  distributions.
  \label{fig:linemapF_0004}}
\end{figure}

In order to extract outstanding characteristics of two-phase
medium, we also perform the simulation of ``one-phase'' 
 isothermal (10K) turbulent flow with the same resolution 
 ($2048^2$ grids). 
This choice of temperature is based on the conventional assumption in
 many isothermal turbulence models of dense molecular clouds
 strongly shielded against UV field \citep[e.g.,][]{padoan01}.
Figure \ref{fig:map_0004} shows the density distributions from a
 snapshot of the one-phase model. 
Filamentary structures, which are commonly observed
in simulations of isothermal turbulent flow, appear in Figure
\ref{fig:map_0004} as well.
Sharp edges of filamentary structures imply that they are
 formed by shock compression.
The synthetic emission maps of [\ion{C}{2}] 158$\mu$m and CO$(1-0)$
lines, which are calculated in a way similar to those of the two-phase
model, are displayed in Figure \ref{fig:linemapF_0004}.
As in the two-phase model, [\ion{C}{2}] and CO$(1-0)$ emission
distributions closely trace the density one.
For the one-phase model, our emissivity maps agree well with previous
 work on, for example, synthetic maps of optically thin $^{13}$CO line
 of the isothermal model of \citet{padoan01}.

\subsection{Analyses of Synthetic Line Intensities}

To extract characteristic properties of the two-phase models from our
calculations, further analysis is required.
We simulate realistic observations of a three-dimensional cube 
 by taking one of the axes of our two-dimensional hydrodynamical 
 calculations as a line of sight.
Defining the $x$ axis as the line of sight and the $x$ component of
the velocity field as the velocity along the line of sight, we can
calculate the intensity of the line emission $I(L)$ as
  \begin{equation}
    I(L) = \frac{1}{4 \pi} \sum_{v_x}\sum_{x}\frac{d\Lambda_{L}}
    {dv_x}(x, v_x) \Delta x \Delta v_x, 
    \label{eq:integ}
  \end{equation}
where $\Lambda_{L}$ is the emission rate per unit volume on the $L$-th
line of sight, $\Delta x$ is the grid size, and 
$\Delta v_x$ is the bin size of the velocity channel, respectively.
Straight summation of $\Lambda_{L}$ is applicable because we
assume all the lines are optical thin.
Hereafter we analyze integrated intensity data obtained by equation
(\ref{eq:integ}). 
Note $\Lambda_{L}$ includes thermal broadening as well as bulk fluid velocity 
(see eq. \ref{eq:profile}), and therefore $I(L)$ correctly reflects
the velocity structure inherent in the model as well.

\subsection{Ratio of High- and Low-Temperature Tracer Lines}
A distinguishing feature of the two-phase turbulence model induced by
thermal instability is the coexistence of the gases of diversely
different thermal conditions.
In this section, we investigate characteristics of the two-phase model by
simultaneous analysis of two kinds of lines that trace either 
high- or low- temperature medium.
In the following sections, we examine intensity ratios between the 
[\ion{C}{2}] line and CO$(1-0)$
line, CO rotational lines, and \ion{C}{1} fine-structure
lines.

\subsubsection{[\ion{C}{2}]-CO(J=1-0) line ratio}

\begin{figure}[htbp]
\epsscale{.80}
\plotone{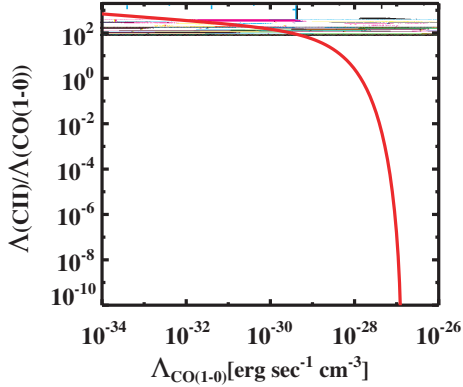}
\caption{Ratio of emissivities per unit volume, 
 $\Lambda_{\mathrm{[C II]}}/\Lambda_{\mathrm{CO(1-0)}}$ as a function of
  $\Lambda_{\mathrm{CO}(1-0)}$. 
  The values of $\Lambda_{\mathrm{[C II]}}/\Lambda_{\mathrm{CO(1-0)}}$ are
  shifted by an arbitrary factor of $3\times 10^{-7}$ for
  clarity of the figure.
  The horizontal line located in the lower half consists of dots
  from the one-phase model, while the upper, slightly bending
  curve consists of those from the two-phase model. The thin solid
 curve is the isobaric contour at the saturation pressure $P_\mathrm{sat}=10^{3.4}$ K cm$^{-3}$.
  \label{fig:CII_co10_p}}
\end{figure}

Figure \ref{fig:CII_co10_p} displays the 
$\Lambda_{[{\rm CII}]} - \Lambda_{\rm CO(1-0)}$ ratio
(in Figure \ref{fig:CII_co10_p} the ratio of
$\Lambda_{[{\rm CII}]}/\Lambda_{\rm CO(1-0)}$ is multiplied by a
constant factor $3\times 10^{-7}$ just for clarity of the figure; 
therefore, in the following arguments, the absolute values of
longitudinal axis are in an arbitrary unit, and only the relative
positions of the points in the diagram have meaning).
The points taken from the one-phase model fall on a straight line of
a constant value of
$\Lambda_{[{\rm CII}]}/\Lambda_{\rm CO(1-0)} \sim 10^{-2}$, while
those from the two-phase model make a slightly bending curve with much
larger values of $\Lambda_{[{\rm CII}]}/\Lambda_{\rm CO(1-0)} \gtrsim
10^2$ in this diagram.
This difference is interpreted as follows:
As we assume that the medium is optically thin, and all the densities
in our calculations are below the critical density for both
[\ion{C}{2}] and CO$(1-0)$ lines ($\approx$ 3000 cm$^{-3}$), the line
  ratio is simply described in terms of $\Lambda$,
  \begin{equation}
  \frac{I_{\rm [C II]}}{I_{\rm CO(1-0)}}
   \approx
   \frac{\Lambda_{\rm [CII]}}
        {\Lambda_{\rm CO(1-0)}} 
   =
     \frac{y({\rm C^{+}}) 
           \langle\sigma v\rangle_{\rm [CII]}(T)}
          {y({\rm CO}) 
           \langle\sigma v\rangle_{\rm CO(J=1)}(T)}. 
  \label{eq:CII-CO}
  \end{equation}
In equation (\ref{eq:CII-CO}) $\langle\sigma v\rangle_{\rm [CII]}$ and
$\langle\sigma v\rangle_{\rm CO(J=1)}$ are determined by the cross
sections for collisional excitations and are dependent only on temperature.
In our calculations the abundances of CO and C$^{+}$ are assumed to
be uniform, and then equation (\ref{eq:CII-CO}) implies that the ratio
$I_{\rm [CII]}/I_{\rm CO(1-0)}$ is a function of temperature.
In our one-phase turbulence model of which temperature is adopted to be
constant (=10K), the temperature-dependence of equation 
(\ref{eq:CII-CO}) drops out and the [\ion{C}{2}] intensity is linearly 
proportional to the CO$(1-0)$ intensity (see Figure
\ref{fig:CII_co10_p}).

On the other hand, the ratio is not constant in the two-phase
turbulence model where temperature strongly varies from place to place.
In this case, the pressure approaches approximately constant within the
whole calculation domain and the points taken from our two-phase model
fall on the isobaric contour of $P= P_\mathrm{sat}$.
We overplot the isobaric contour with $P= P_\mathrm{sat}$ with a solid
line in Figure \ref{fig:CII_co10_p}, which closely trace the dots from
our two-phase simulations.
Under constant pressure, increasing density indicates decreasing
temperature.
In the low density (or high temperature $T\gtrsim$ 1000K) limit the
cooling rate per unit volume increases with density as $n_{\rm
  H}^{2}$, and the collisional cross sections $\langle\sigma v\rangle$
are similar for both C$^{+}$ and CO in such a high-temperature regime.
Thus $\Lambda_\mathrm{CO(1-0)}$ is approximately proportional to
$\Lambda_\mathrm{[C II]}$.
On the other hand, as the temperature decreases below $T\lesssim
100{\rm K}$, $\langle\sigma v\rangle_{\rm [CII]}$ decreases with $T$
much faster than $\langle\sigma v\rangle_{\rm CO(J=1)}$ because of the
exponential decrease in the number of particles that can excite the
[\ion{C}{2}] line (92K). 
Simultaneous increase in density with a decrease in
temperature brings the number of CO molecules in the level $J=1$ 
close to the LTE limit.
Because of the slower decrease in $\langle\sigma v\rangle_{\rm CO(J=1)}$
with temperature compared with $\langle\sigma v\rangle_{\rm [CII]}$, 
and the saturation of the $J=1$ level population owing to LTE, there
is a sharp decline of the isobaric contour toward 
$\Lambda_\mathrm{CO(1-0)} \approx  10^{-27}$ erg
sec$^{-1}$ cm$^{-3}$.
This explains the behavior of the isobaric line in Figure
\ref{fig:CII_co10_p}, which turns downward with increasing density.

\subsubsection{CO rotational line ratios} \label{sect:COratio}
\begin{figure}[htbp]
\epsscale{1.0}
\plotone{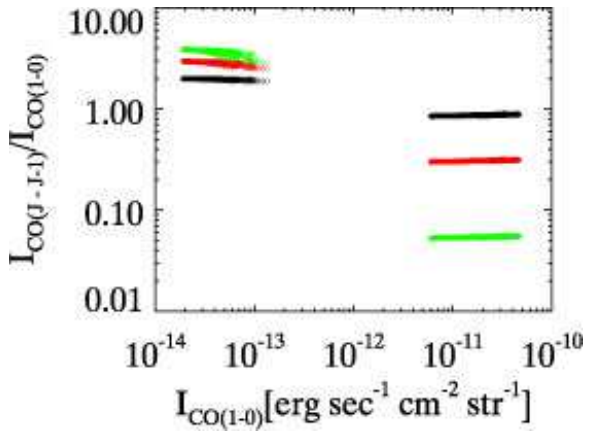}
\caption{The ratios of intensities of a CO high ($J=2,3,4$) excitation 
line with the CO$(1-0)$ line. 
  There are two sequences of plots corresponding to the two-phase and
  the one-phase model, respectively.
  The green dots are $R_{43/10}$, the red are $R_{32/10}$, and the
  black are $R_{21/10}$.
  The left sequence is the two-phase flow
  data, and the right one is the one-phase flow data.
  In the two-phase model, the line ratio $R_{J,(J-1)/10}$ increases
  with $J$, while, in the one-phase model, its dependence on $J$ is
  opposite for $J\le$ 4.
 \label{fig:CO_ratio}}
\end{figure}

In this subsection, we consider ratios between intensities of CO$(1-0)$ 
and higher $J$ lines.
Since the excitation energies of rotational levels from the ground
state $E_{J}/k_B=2.75{\rm K} J(J+1)$ increase with $J$, 
higher $J$ lines trace higher temperature gas.
In Figure \ref{fig:CO_ratio} we show the line intensity ratios $R_{J,
  (J-1)/10} \equiv I_{\mathrm{CO}(J-(J-1))}/I_{\mathrm{CO}(1-0)}$ 
for both two-phase and one-phase models.
The dependence of line ratios on rotational quantum number $J$ is 
surprisingly different for these two turbulent flow models.
In the one-phase model, the line ratios $R_{J,(J-1)/10}$ decrease 
with increasing $J$, while this trend reverses in the two-phase model
up to $J\lesssim 5$: in the one-phase model, although the CO$(2-1)$ line 
is almost as strong as CO$(1-0)$ line ($R_{21/10} \lesssim 1$), 
higher $J$ lines become progressively weaker.
On the other hand, in the two-phase model, the line ratio 
$R_{21/10}$ is as large as 2 and $R_{32/10}$ and $R_{43/10}$ become
even larger.

This can be explained by the following argument:
Since density is lower or at most marginally 
higher than the critical density 
($n_{\rm cr} \sim 3000{\rm cm^{-3}}$) in our snapshots,
here we consider the low-density limit.
In other words, we discuss the cases where the radiative 
decay from each level dominates all the collisional transitions 
both into and from this level, except the collisional excitation 
from the ground state.
Under this assumption, the equations of the detailed balance for the
optically thin lines can be approximately described as
  \begin{equation}
    A_{10}n_{1}= \sum_{J \geq 1} C_{0J} n_{0}
    \label{eq:balance1}
  \end{equation}
for the ground level ($J = 0$) and 
  \begin{equation}
    A_{J+1,J}n_{J+1}+C_{0J}n_{0} = A_{J,J-1}n_{J},
    \label{eq:balance2}
  \end{equation}
for levels $J \geq 1$. 
The collisional excitation rate $C_{ij}$ is written 
as in equation (\ref{eq:Cij}).
From these equations, we obtain
  \begin{equation}
    A_{J,J-1}n_{J} = \sum_{J^{\prime} \geq J} C_{0J^{\prime}} n_{0}.
    \label{eq:radrate}
  \end{equation}
Equation (\ref{eq:radrate}) implies that all the CO molecules that
are collisionally excited to the state $J$ cascade down to lower
levels by spontaneous emission with $\Delta J=1$.
Then the downward transition rate from the state $J$ by spontaneous
emission is equal to the sum of the collisional excitation rate from the
ground state to all the states with $J^{\prime} \geq J$.
Using equations (\ref{eq:coolingrate}) and (\ref{eq:radrate}), the
ratio between the adjacent lines $(J+1)\rightarrow J$ and
$J\rightarrow (J-1)$ is 
  \begin{eqnarray}
   R_{(J+1),J/J,(J-1)} &=& \frac{E_{J+1,J}}{E_{J,J-1}} 
     \frac{\sum_{J^{\prime} \geq J+1}\gamma_{0 J^{\prime}}}
          {\sum_{J^{\prime} \geq J} \gamma_{0 J^{\prime}}}
     \label{eq:ratio1} \\
   &=& \frac{E_{J+1,J}}{E_{J,J-1}} \frac{G_{J+1}}{1+G_{J+1}},
  \end{eqnarray}
where $G_{J+1}$ is defined as
  \begin{equation}
    G_{J+1}=\sum_{J^{\prime} \geq J+1}\frac{\gamma_{J^{\prime}0}}{\gamma_{J0}}
    \frac{g_{J^{\prime}}}{g_{J}} 
    \exp{\left( -\frac{E_{J^{\prime},J}}{k_BT}\right)}. \label{eq:ratioG}
  \end{equation}
In the above, we have used the relation 
\begin{equation}
\gamma_{0J}=\gamma_{J0} \frac{g_{J}}{g_{0}} 
\exp{\left( -\frac{E_{J,0}} {k_BT}\right)},
\label{eq:gamma}
\end{equation} 
where $g_{J}$ is the statistical weight of the level $J$.
\begin{figure}[htbp]
\epsscale{.80}
\plotone{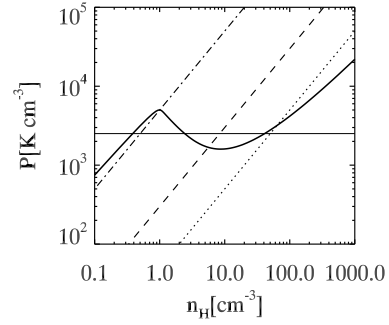}
\caption{Thermal equilibrium curve ($\Gamma=\Lambda$, thick solid
  curve), and isothermal lines of $T=$ 5000K (dot-dashed), $T=$ 300K
  (dashed), and $T=$ 50K (dotted : the lowest temperature of the
  snapshot of two-phase model) in the pressure-density diagram.
  The saturation pressure $P=P_\mathrm{sat}$ is plotted with the thin
  solid line.
  The temperature range where bi-stability is possible is
  limited to 300K$\lesssim T \lesssim$5000K.
 \label{fig:th_eqF}}
\end{figure}

In the following, we examine the line intensity ratios using the set
of equations (\ref{eq:ratio1})-(\ref{eq:ratioG}) in the low and high
temperature limits.
In the low temperature limit where $T \ll E_{J+1,J}/k_B$, $G_{J+1}$
can be approximated with its first term,
  \begin{equation}
    R_{(J+1),J/J,(J-1)} \simeq \frac{E_{J+1,J}}{E_{J,J-1}} 
    \frac{\gamma_{J+1,0}}{\gamma_{J,0}}
    \frac{g_{J+1}}{g_{J}} \exp\left(-\frac{E_{J+1,J}}{k_BT}\right).
    \label{eq:ratiolow}
  \end{equation}
Equation (\ref{eq:ratiolow}) shows that when the temperature is lower
than the excitation energy $E_{J+1,J}/k_B$, the line ratio
$R_{(J+1),J/J,(J-1)}$ is exponentially reduced.
In our one-phase model with $T$=10K, the excitation energy
$E_{J,J-1}/k_B$ is higher than the kinetic temperature for high $J$
($J\ge 2$) lines and the line ratio
$R_{J,J-1}/R_{10}=\prod_{J^{\prime}=1}^{J} R_{J^{\prime}+1,
  J^{\prime}/J^{\prime},J^{\prime}-1}$ is significantly smaller than
unity. 
The only exception is $R_{21/10}\sim 1$ since $E_{1,0}/k_B=11$K is 
comparable to the kinetic temperature.
On the other hand, in the high-temperature limit where $T\gg E_{J+1,
  J}/k_B$, $G_{J+1} \gg 1$ since numerous terms in the summation
of $G_{J+1}$ (eq. [\ref{eq:ratioG}]) are order of unity. 
Then,
  \begin{equation} 
    R_{(J+1),J/J,(J-1)} \simeq \frac{E_{J+1,J}}{E_{J,J-1}}. 
    \label{eq:ratiohigh}
  \end{equation}
In our two-phase model, high-temperature regions where high
$J$ lines ($T \ga E_{21}/k_B$) can be collisionally excited are
present.
In the snapshot shown in Figure \ref{fig:linemapF_0012}, indeed, the
temperature ranges from 50K to 3700K which is higher than CO
excitation energies up to $J\le$ 4 ($E_J = 55$ K) in most of the
volume (see Fig. \ref{fig:th_eqF}).  
The excitation energies up to $J\le 4$ are smaller than 50K.
Therefore equation (\ref{eq:ratiohigh}) is applicable to the whole
region and $R_{J, J-1/10}$ becomes order of unity.

\subsubsection{[\ion{C}{1}] line ratio}
\begin{figure}[bthp]
\epsscale{.80}
\plotone{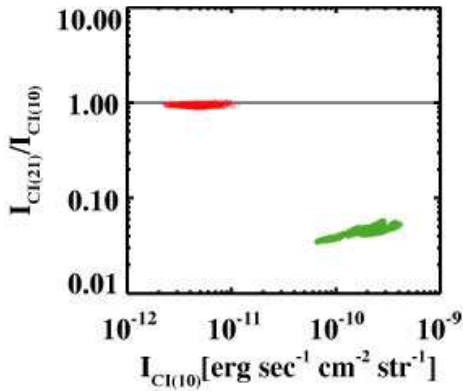}
\caption{The ratios of \ion{C}{1} fine-structure-line intensities
  $I(^3P_2-^3P_1)/I(^3P_1-^3P_0)$ as a function of $I(^3P_1-^3P_0)$.
  The two sequences correspond to the one- and two-phase models.
  The ratios of high- and low-transition-energy lines of C$^0$
  resemble those of CO molecules: in the two-phase model, the ratio is
  close to unity, while in the one-phase model the ratio is smaller
  than 0.1.
  See discussion in \S\ref{sect:COratio}.
 \label{fig:CI_ratio}}
\end{figure}

In the transition region between the warm atomic region of low density
and the cold molecular region of high density, carbon is partly in a
neutral atomic form C$^0$.
In this section, we calculate [\ion{C}{1}] fine-structure lines.
In the temperature and density range under consideration, important lines
are the forbidden transitions among the triplet $^{3}P_{2}$, $^{3}P_{1}$ 
and $^{3}P_{0}$ in the ground electronic state.
In Figure \ref{fig:CI_ratio}, we show the ratio of
$^{3}P_{1}-^{3}P_{0}$ (24K) and $^{3}P_{2}-^{3}P_{1}$ (39K) lines.
Similar to CO rotational line ratios, the [\ion{C}{1}] line ratio
becomes close to unity in the two-phase model, while it is
significantly smaller than unity in the one-phase model.
This behavior can be understood in a similar way to the argument 
for CO rotational lines (\S\ref{sect:COratio}).
Again, we consider the low-density limit.
In the following we call levels $^{3}P_{0}$, $^{3}P_{1}$, and $^{3}P_{2}$
 levels 0, 1, and 2, respectively.
Since the radiative transition from level 2 to 0 
is negligible compared with that from level 2 to 1, 
the equations for statistical balance read
  \begin{equation}
  A_{21}n_{2}+C_{01}n_{0} = A_{10} n_{1}
  \end{equation}
for level 1, and 
  \begin{equation}
  C_{02}n_{0} = A_{21} n_{2}
  \end{equation}
for level 2.
From these equations and relations (\ref{eq:Cij}) and (\ref{eq:gamma}),
the line ratio is given by  
\begin{eqnarray}
R_{21/10}&=&\frac{E_{2,1}A_{21}n_{2}}{E_{1,0} A_{10}n_{1}}\\ 
&=& \frac{E_{2,1}}{E_{1,0}} \frac{\gamma_{02}}{\gamma_{01}+\gamma_{02}}\\
&=& 1.625 \frac{G}{1+G}          
\end{eqnarray}
where 
\begin{eqnarray}
G &=& \frac{\gamma_{20} g_{2}}{\gamma_{10} g_{1}}  
\exp{\left(-\frac{E_{2,1}}{k_{\rm B} T} \right)} \nonumber \\
&=& 0.96 \left( \frac{T}{100{\rm K}} \right)^{0.12}   
\exp{\left(-\frac{39 {\rm K}}{T} \right)},
\end{eqnarray}
where we consider only collision with neutral hydrogen as for the
excitation partner for numerical evaluation.
The above expression clearly expresses that the line ratio becomes close to 
unity for $T \ga 39{\rm K}$ (as in the two-phase model),
 while for lower temperature (as in the one-phase model of $T=10$K) 
it decreases exponentially.

\subsubsection{Summary of the Line-ratio Analyses}
We have demonstrated in this subsection that the intensity ratio
$R_{h/l}$ of lines of high-transition energy ($\Delta E_{h}/k_B
\approx$ 100K) and low-transition energy ($\Delta E_{l}/k_B \approx$
10K) is as high as unity in the two-phase model,  while $R_{h/l}$ is
much smaller than unity in the 10K isothermal model.
This feature is quite robust and is independent of the species of
emitting particles (atomic carbon or CO molecule) and the
nature of transitions (fine-structure transitions or rotational
transitions).
Discussion in \S\ref{sect:COratio} explains the trend in line ratios in
terms of the level population in low-density gas.
The level populations are determined by collisional excitation and
radiative de-excitation by spontaneous emission.
In this case the number of particles that have enough energy for level
excitation is proportional to $\exp(-\Delta E/k_BT)$, and then the
level population reflects the ratio of transition energy and kinetic
energy, $\Delta E/k_BT$. 
In other words, our results show that simultaneous observation of both
high- and low-energy lines from the diffuse atomic/molecular
transition region will reveal the thermal structures of the gas.
If the high line ratio is observed in such regions in future, 
we can confirm the existence of two-phase turbulent flows.
In the following section, we discuss the implications of our results
for the CO line ratio in more detail.

\section{Discussion}

\subsection{Possible CO abundance variation in the warm phase}
\label{sect:COabund}

We have assumed uniform abundance distribution both temporally and 
spatially for all the species.
This assumption appears too artificial if we consider complex chemical 
reactions in the temperature range $T\lesssim 10^3$K.
Actual chemical abundances in diffuse clouds will evolve both in space and
time, reflecting the environments of the clouds and their evolutionary
history, that is,  the development of the multi-phase medium.
In particular, whether CO is present in such a diffuse environment is
quite uncertain.
Since line intensity is proportional to the abundance of emitting
species in an optically thin medium, the lack of CO in the warm gas
would reduce the CO line intensity.
In Figure \ref{fig:linemapF_0012}a, one can see that warm ($T\sim
1000$K) gas occupies a large fraction of volume in the two-phase
medium.
In evaluating the CO line intensity by integrating the emissivity 
along the line of sight (or the $x$ axis in our analysis), 
we need to take into account the possible deficiency of CO molecules 
in the warm phase.

To assess the extent of this effect on our results, we perform the 
following simple experiments:
Although formation and destruction rates of CO molecules are dependent 
on various thermal factors, e.g., density and temperature,
the degree of depletion onto dust grains and so on in a complex
fashion, it is plausible to assume that the abundance of CO, 
$y(\mathrm{CO})$ increases with decreasing temperature and increasing
density.
In our two-phase model, the whole region is approximately in pressure
 balance. 
Thus higher density means lower temperature,  and 
 CO abundance can be regarded as a function only of temperature, 
 as long as chemical reactions of CO are in equilibrium. 
As the simplest example, we assume that CO is completely destroyed
above a threshold temperature $T_\mathrm{crit}$.
Namely, the CO abundance $y(\mathrm{CO})$ is taken as a 
step function of temperature $T$;
  \begin{equation}
    y(\mathrm{CO}) = \cases{ 
      3\times 10^{-7} & $T\le T_\mathrm{crit}$, \cr
      0 & $T > T_\mathrm{crit}$. \cr }
  \end{equation}
We take $T_\mathrm{crit}$ as a free parameter and
examine the effect of different values $T_\mathrm{crit}$ on the
line ratios.
Figure \ref{fig:th_eqF} shows that the medium is
thermally unstable in the range of 300K $\lesssim T \lesssim$
5000K. 
Therefore the spread of temperature grows as large as this range in
our two-phase flow calculations (50K $\lesssim T \lesssim$ 3,700K).
We study four cases of $T_\mathrm{crit}$ = 100K, 400K,
1000K, and 3000K, which correspond to $n_H=40$, 10, 4, and 1.3 cm$^{-3}$,
respectively, under the saturation pressure $P_\mathrm{sat}=10^{3.4}$ K
cm$^{-3}$.
We repeat the same analyses as in \S 3 for each of $T_\mathrm{crit}$.

\begin{figure}[bthp]
\epsscale{1.0}
\plotone{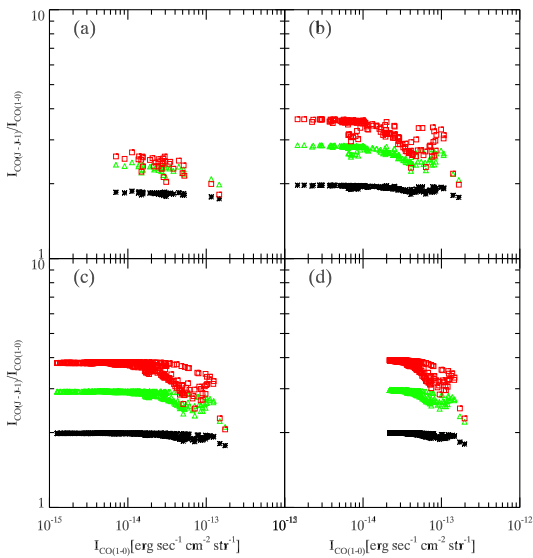}
\caption{CO rotational line ratios with different values of 
threshold temperature $T_\mathrm{crit}$,
  above which CO abundance is set to zero.
  Dots of each color correspond to the same transition $J$ as those
  displayed in Figure \ref{fig:CO_ratio}.
  The threshold temperatures are (a) $T_\mathrm{crit}=100$K, 
  (b) 400K, (c) 1000K, and (d) 3000K.
  If the pressure is at the saturation value 
  $P_\mathrm{sat}=10^{3.4}$ K cm$^{-3}$,
  the equivalent threshold densities are (a) 40 cm$^{-3}$, 
 (b) 10 cm$^{-3}$, (c)4 cm$^{-3}$, and (d) 1.4cm$^{-3}$, respectively.
  One can observe that as $T_\mathrm{crit}$ increases, the value of 
 $I(\mathrm{CO(1-0)})$ increases.
  This is because emission from the warm gas is added to the summation 
  along the line of sight.
  Also the values of the ratio do not change significantly,
  because, even in the coldest regions, the energy difference of the levels 
  $\Delta E/k_BT < 1 $ in the two-phase model 
 (see discussion in \S\ref{sect:COratio}).
 \label{fig:CO_ratio_tcut}}
\end{figure}

In Figure \ref{fig:CO_ratio_tcut}, we show the CO($J-(J-1)$)/CO$(1-0)$
line ratios for different values of $T_\mathrm{crit}$.
With a decrease in $T_\mathrm{crit}$, the volume occupied by the warm
phase (where $y(\mathrm{CO})$ is set to be 0) drastically increases. 
Radiation is emitted only from surviving cold clumps.
In spite of a drastic increase in volume of the warm phase, the line ratios 
do not change significantly (Figure \ref{fig:CO_ratio_tcut}).
This result is supported by our discussion in \S\ref{sect:COratio}
that the line ratio of CO rotational transitions depends only on the ratio of 
the excitation energy of the line $\Delta E$, and the kinetic
temperature of the gas $k_B T$, as long as density is low
($n<n_\mathrm{crit}$).
As depicted by the isothermal lines in Figure \ref{fig:th_eqF}, after
the system settles onto a constant pressure around the saturation
pressure $P_\mathrm{sat}$, the lowest temperature of the two-phase
model is as high as 50K in our calculation. 
This lowest temperature assures $E_J/k_B T <1$ 
for $J\lesssim 4$ ($T_J=$ 55K), so that even in the coldest region in our 
two-phase model the line ratios $R_{J,(J-1)/10}$ exceed (or are close to) 
unity.
Since CO molecules are preferentially formed in the dense
strongly shielded clumps (see also discussion in this section below), 
the temperature inside the CNM of the two-phase ISM can be lower 
than our results owing to the resultant CO cooling.
In fact, according to \citet{wolfire95}, the thermal equilibrium 
temperature in CNM is as low as $\sim$ 20K.
The condition for high line ratio $E_J/k_B T < 1$ becomes
more difficult to be satisfied in colder CNM: the line ratio is
accordingly lowered.
Thus, the value of $R_{J,(J-1)/10}$ of higher $J$ is reduced, while
$R_{21/10}$ stays the same as our results except in  
strongly shielded molecular clouds of 10K owing to the very low
excitation energy only 16.5K. 
In this case we should use the line ratio $R_{J,(J-1)/10}$ of low $J$
lines to study the thermal structure of the CNM.

The increase in the threshold temperature $T_\mathrm{crit}$ 
is accompanied by an increase in the average temperature $T$ in 
the emission region. 
Since the line ratio $R_{J,(J-1)/10}$ grows larger 
as the average temperature $T$ increases (\S \ref{sect:COratio}), 
the shift of the threshold temperature $T_\mathrm{crit}$ 
does not alter the trend toward the averaged $R_{J,(J-1)/10} > 1$ as
long as $\Delta E/k_BT < 1$.
Note that our simple experiments include two extreme cases of high- and
low- $y(\mathrm{CO})$ values in the warm phase.
Thus we conclude that, as long as we are concerned with the line-ratio 
analysis presented in this article, the result is not significantly affected
by complicated chemical evolution.
In other words, the line ratio, not the intensity of each line, is a 
good observational tool for understanding the thermal structure of
turbulent interstellar medium.

Since the chain reaction for formation of CO molecules starts with a
reaction involving H$_2$, the presence of H$_2$ is a necessary condition
for CO formation. 
H$_2$ molecules are formed via association of two hydrogen atoms on 
the surface of dust grains.
The formation timescale of H$_2$ molecules is described as
$\tau_{\mathrm{H}_2}=1.67\times 10^9n_{\rm H}^{-1}$ years 
(\citealt{spitzer78}).
The formation timescale becomes on the order of $10^6$ years
in the dense cold clumps, where $n_H \ga 10^3$ cm$^{-3}$.
Then if the dense clumps survive as long as $10^6$ years, CO molecules 
can form inside.
There are some arguments supporting the existence of 
CO molecules in tenuous warm ISM. 
Recently, \citet{inoue06} emphasized the importance of gas flows
(i.e. evaporation) from cold phase to warm phase.
If molecules are abundant inside the cold clouds, they will be
dispersed into the warm phase.
The efficient evaporation of CO from cold clumps
enables the supply of CO molecules to the warm gas 
on a relatively short timescale, although the chemical timescale is 
much longer than other relevant timescales in the warm phase.
Of course, if the UV radiation is as intense as in average diffuse 
ISM, such CO molecules would be photodissociated immediately.
\citet{hennebelle06} proposed an alternative model for cold molecular
clouds that possess warm gas deep inside.
In this picture, CO molecules can avoid photodissociation within such
warm neutral medium (WNM) because external UV radiation is blocked by
the surrounding cold molecular gas.
In this scenario, the condition $\Delta E/k_BT <1$ is 
satisfied for CO molecules in WNM.

Here we discuss two possible evolutionary scenarios for CO abundance 
 in diffuse ISM. 
First, we consider the case of relatively strong interstellar UV
radiation, stronger than the adopted value ($G_0 = 1$ in our article).
If there were a large number of FUV photons, CO molecules
would be easily photodissociated.
The shielding of UV photons by atomic and molecular hydrogen 
 and dust grains would result in selective destruction of 
 CO molecules in the tenuous warm medium, 
 but CO molecules in dense, cold clumps, surrounded by a warm medium,
 would be protected against UV penetration.
Another possibility is that a strong UV field might thermally stabilize 
the medium at a single warm phase \citep{wolfire95}.
In this case, however, CO molecules would not be able to survive 
photodissociation and CO line emission would be very
weak, even though there are some turbulent flows.
The survey observations of molecular clouds in the Galactic plane
suggest that the number of regions of high-line ratio
$R_{21/10}\gtrsim 1.0$ is small \citep{sakamoto97,oka98}.
These results might be interpreted as a signature of the lack of
initial CO molecules by photodissociation.
If this is the case, more refractory molecules must be used as 
tracers of the two-phase medium.
In order to examine what transitions can be better tracers,
it is necessary to include greater detail of chemical evolution in our
 hydrodynamical simulations, as well as of the evolution of the main
 cooling and heating sources.
\footnote{Equation (\ref{eq:fitting}) describes the cooling rate of
  [\ion{C}{2}] 158 $\mu$m line, the major coolant in tenuous ISM with
  which we are concerned in this article.}
Second, we consider the case of weak UV irradiation as we have
  considered in this article.
In our two-phase models, the temperatures do not exceed $T\approx$
  5000K, and CO molecules once formed are not subject to the
  collisional dissociation. 
In this case, our experiments in this subsection are straightforwardly
  applicable and the ratio of rotational transitions of CO molecules
  is a good probe of thermal structure in the turbulent ISM.

\subsection{Comparison with Observational Results}
\subsubsection{Related Observations of Diffuse Clouds}

The effective cooling function we adopted (equation \ref{eq:fitting}) 
 was calculated under the condition of small visual extinction, 
 $A_V\lesssim 0.1$.
This environment corresponds to the periphery of molecular clouds where 
the composition of carbon changes from atomic to molecular form.
\citet{sakamoto03} found peculiar structures in the position-velocity
 (P-V) diagram around $A_V \lesssim 1$ in their observations of
 CO$(1-0)$ in Taurus HCL2, which spans over a wide range of $A_V$. 
Tiny clumps with sharp velocity gradients of radius $R\approx 0.05$pc
suddenly appeared when $A_V$ fell below unity in all of three observed
strips. 
They speculated that these structures were cold dense clumps formed by 
thermal instability since this instability under mild UV heating
occurred approximately around $A_V \sim 1$.
In Figure \ref{fig:co_pv} we display the synthetic 
P-V diagram of CO$(1-0)$ line from the snapshots of the simulations
at 3 Myr.
The two-phase model shows that velocity of clumpy 
structures spreads over a wide range ($\Delta v_r \approx$ 5 km sec$^{-1}$).
Indeed, our synthetic P-V diagram of the two-phase model looks
quite similar to that of \citet{sakamoto03} at $A_V\lesssim $1.
Since our simulations do not consistently include the spatial variation 
of UV shielding in modeling periphery of molecular clouds, 
direct comparison with these observational results should be done with caution.
This similarity is, however, quite encouraging for confirming our
two-phase model driven by UV photoelectric heating. 

IRAM observations of starless molecular clouds also suggest 
the possibility that tiny gas clumps are filling the beam 
\citep{falgarone98}.
Furthermore, in more diffuse regions in the Galactic plane, ISO
observations of H$_2$ rotational lines suggest
the existence of collisionally heated warm gas ($T\lesssim 10^3$K)
within the cold diffuse ISM \citep{falgarone05}.
Although both of the observed regions are denser than our
calculations, their structures might be roughly consistent with the
two-phase medium models generated by UV photoelectric
heating \citep{koyama02,audit05} or dissipative heating of Alfv\'en
waves \citep{hennebelle06}. 
We can clarify the nature of the multi-phase ISM by making comparisons 
with line-emission calculations like ours and such detailed observations 
of diffuse ISM as described above.

\subsubsection{Comments on observations of the CO line ratio $R_{21/10}$}

Despite our conclusion that the two-phase medium has a high line-ratio 
$R_{J, J-1/10} =$ \par $I_{\mathrm{CO}(J-(J-1))}/I_{\mathrm{CO}(1-0)} \gtrsim$ 1.0, 
observations have found that regions of high line-ratio
$R_{21/10} =$ $I_{\mathrm{CO}(2-1)}/I_{\mathrm{CO}(1-0)}$ $\gtrsim$
1.0 are of limited fraction \citep{sakamoto97, oka98} and usually only
appear close to HII regions.
Our results of high $R_{21/10}$ values are in part due to the
assumption of small optical thickness ($\tau\ll$ 1).
If we included the effect of finite optical thickness, 
the line ratio $R_{21/10}$ averaged over the simulation box would 
become smaller.
However, as for the peripheral regions, which are optically thin, our
treatment is still valid and our conclusions should not be altered.
Discrepancy between existing observations and our prediction is probably 
due to the fact that observations tend to be strongly biased 
to clouds with high CO intensity, thus having higher average density 
compared with the transition regions we discussed.
To observationally examine the atomic/molecular transition regions, 
we must take care not to include both high and low CO intensity
regions in a single observing beam in resolving the spatial
gradient of $A_V$.
For this purpose, observations not only with high angular resolution
but also with high sensitivity are needed for achieving high S/N
observation of faint transition regions, avoiding contamination of
high intensity regions.
We hope that future instruments will provide a good information 
for the two-phase scenario of ISM turbulence.

Here we discusss some of the characteristics of the cold clumps
analyzed in this paper and future observation.
The typical size of the cold clumps is determined by the maximum growth
wavelength in thermally unstable medium $\lambda_\mathrm{MGR}\approx
\sqrt{\lambda_\mathrm{F}\lambda_\mathrm{cool}}$ \citep{field65}, where
$\lambda_\mathrm{F}=(KT/{n_{\rm H}}^2\Lambda)^{1/2}$ is the Field length, and
$\lambda_\mathrm{cool} = c_s \cdot (3/2)k_{\rm B}T/({n_{\rm H}}\Lambda)=
c_s \tau_\mathrm{cool}$ is the cooling length,
respectively.
Numerically these values are
 \begin{mathletters}
   \begin{eqnarray}
     \lambda_\mathrm{F} &=& 7.6 \times 10
       \left( \frac{T}{\mathrm{300K}}\right)^{3/2}
       \left( \frac{n_{\rm H}\Lambda}{10^{-25} {\rm erg~sec^{-1}}}\right)^{-1/2} \nonumber \\
       & &
       \left( \frac{n_{\rm H}}{\mathrm{100 cm}^{-3}} \right)^{-1/2} \mathrm{AU,}\\
     \lambda_\mathrm{cool} &=& 3.6 \times 10^3
       \left( \frac{T}{\mathrm{300K}}\right)^{3/2}
       \left( \frac{n_{\rm H}\Lambda}{10^{-25}{\rm erg~sec^{-1}}} \right)^{-1}
     \mathrm{AU,} \nonumber \\
     & &  \\
     \lambda_\mathrm{MGR} &=& 5.3 \times 10^2
       \left( \frac{T}{\mathrm{300K}}\right)^{3/2}
       \left( \frac{n_{\rm H}\Lambda}{10^{-25}{\rm erg~sec^{-1}}}\right)^{-3/4}
       \nonumber \\
     & &
       \left( \frac{n_{\rm H}}{\mathrm{100 cm}^{-3}}\right)^{-1/4} \mathrm{AU,}
   \end{eqnarray}
 \end{mathletters}
where ${n_{\rm H}\Lambda}$ is cooling rate per particle, and
the normalizations are taken from typical values in our calculation.
If tiny cold clumps reside at the distance of nearby molecular clouds
$D=$ 140pc, the angular size of a clump of radius
$\lambda_\mathrm{MGR}=5.3\times 10^2$AU is $\sim 3.7^{\prime\prime}$.
From Figures \ref{fig:CO_ratio}, \ref{fig:CI_ratio} and
equation (\ref{eq:integ}) intensities of CO($1-0$) and [\ion{C}{1}]
lines of the two-phase medium are $I(\mathrm{min}) \gtrsim 10^{-13}
(\Delta x/2.4 \mathrm{pc})$ erg sec$^{-1}$ cm$^{-2}$ str$^{-1}$ in our
calculations (where $\Delta x$ is the total length scale of the
two-phase medium along the line of sight).
This intensity value corresponds to the brightness temperature $T_b =
2.46\times 10^{-4}$K ($\Delta x/2.4\mathrm{pc}$)K, and is too weak to
be detected with ALMA telescope unless the line-of-sight length of the
cloud is sufficiently large (ALMA Sensitivity Calculator)\footnote{
  URL: {\tt http://www.eso.org/projects/alma/science/bin/sensitivity.html}},
even though its high angular resolution is sufficient for resolving
each tiny clump.
This might suggest that we should target near-by clouds such as some
of the high latitude clouds \citep[e.g.,][]{magnani1985}.
In addition to the weak average intensity, the volume filling factor
of the tiny CNM clumps of which intensity dominates, is much less
than unity (Fig. \ref{fig:linemapF_0012}).
Taking into account of these difficulties, we should be careful in the
interpretation of the observations for the tiny CNM clumps in the
diffuse two-phase ISM even with ALMA.

\begin{figure}[bthp]
\epsscale{1.0}
\plotone{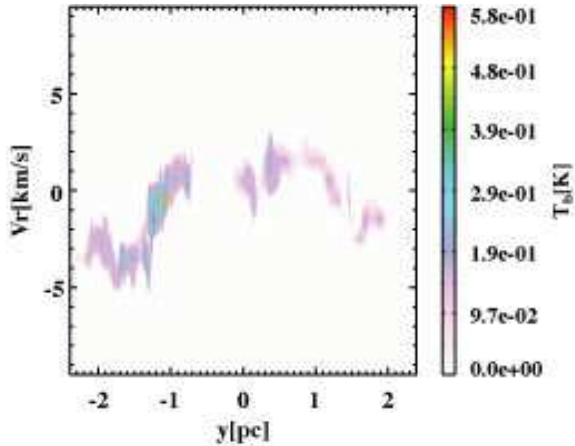}
\caption{Synthetic position-velocity diagram of the CO$(1-0)$ line at
  $\sim$ 3Myr snapshots of the two-phase model.
  Tiny structures spread with a velocity dispersion of $\Delta
  v_r\sim 5$ km/s.
  \label{fig:co_pv}}
\end{figure}
On the other hand, Figure \ref{fig:co_pv} shows that the tiny
clump-like structures are assembled into larger structures of
$\lesssim$ 0.1 pc when integrated along a line of sight in the P-V
diagram.
The flux density of such assembly is
\begin{eqnarray}
 F_{\nu} &=& 0.38
           \left( \frac{I}{1\times 10^{-13} \mathrm{erg} ~
            \mathrm{sec}^{-1} ~ \mathrm{cm}^{-2} ~ \mathrm{str}^{-1}} \right)
           \left( \frac{\Delta v_x}{1 \mathrm{km} ~
             \mathrm{sec}^{-1}}  \right)^{-1}
           \left( \frac{\Delta x}{2.4\mathrm{pc}}\right) \nonumber \\
         &&\times
           \left( \frac{R}{0.05 \mathrm{pc}}\right)^2
           \left( \frac{D}{140 \mathrm{pc}} \right)^{-2} ~ ~
           \mathrm{mJy},  \label{eq:flux}
\end{eqnarray}
which might be observable with a future telescope equipped with highly
sensitive receivers.
For better evaluation of the intensity, we need to study the global
structure of the atomic/molecular transition region which supposedly
consists of the tiny clump-like structures appeared in our
simulations.
Such a global modelling is beyond the scope of this paper, but should
be included in future simulations.

When future highly sensitive multi-$J$ line observations of ``peculiar''
structures with typical size $\sim$ 0.1pc detected in the
strip-scanned CO($1-0$) data (\citealt{sakamoto03}; see \S 4.2.1) are
available, they will greatly improve our understandings of their origin
and identity.
In those rims of molecular clouds with $A_V\lesssim 1$, even smaller
structures might be found with future deep observations.
Small ($\lesssim$ 0.1 pc) structures with high-line ratio in imaging
observations would be the best verification of the two-phase ISM
turbulence models and of the formation scenarios of molecular clouds based on
the multi-phase turbulence models \citep[e.g.,][]{koyama00,bergin04}.
Detailed observations using forthcoming powerful instruments 
will reveal the thermal structure in diffuse interstellar medium, as
well as in dense molecular clouds.
We hope our calculation of the carbon line
ratio would be a prototype to study the nature of ISM using
forthcoming high quality observational data.

\section{Summary and Conclusion}

We examined observational characteristics of the two-phase 
turbulence model by combining hydrodynamical simulations and 
radiation calculations.
We modeled the diffuse interstellar medium irradiated by UV radiation
of $G_0=1$, attenuated with an average visual extinction of $A_V\sim
0.1$.
In our model settings, the two-phase structures are formed by
photoelectric heating, and therefore our model corresponds to the
atomic/molecular transition region at the periphery of dense molecular
clouds.
In order to examine such transition regions, we calculated line
intensities based on nonLTE level populations for
three species of carbon, C$^{+}$, C$^{0}$, and CO.
By comparing two line intensities of high- and low- temperature
tracers, we found that the line ratio $R_{h/l}$ distinctly reflects
the thermal structure of the medium.
The results of our analyses showed that, in the two-phase model, 
higher transition energy lines can be stronger than those of lower
energy as long as $\Delta E/k_BT <1$.
Our argument on the level population in low-density limit
($n<n_\mathrm{crit}$) demonstrated that the line ratio of high- and
low- temperature tracers are determined by the ratio of transition
energy and gas kinetic energy, $\Delta E/k_BT$.
In addition to uniform abundance calculations, we also analyzed 
cases where CO is dissociated above a threshold temperature, or 
equivalently, below a threshold density. 
Our analyses showed that abundance distribution of CO molecules in the
warm phase does not significantly affect the high line ratio that
characterize the two-phase model, as long as CO molecules are not
completely photodissociated.
This is because, even in the coldest regions in 
our two-phase model, the temperature is as high as 50K 
($\approx E_4/k_B$, the energy level of $J=4$ of the CO rotational
transition),  and satisfies $\Delta E/k_BT <1$.
By seeking these features in the diffuse atomic/molecular transition regions,
 we will be able to confirm the existence of two-phase interstellar
 turbulence by using the high angular resolution and high sensitivity
 of future instruments.

\acknowledgments
We greatly appreciate K. Tomisaka, K. Wada, and S. Miyama for their
useful comments on the performance of our analyses, and S. Sakamoto 
for discussions about observations of diffuse interstellar clouds.
We also thank K. Sakamoto for his helpful comments on interferometer
observations. 
This research was supported in part by Grant-in-Aid by the Ministry 
 of Education, Science, and Culture of Japan 
 (16204012, 18026008, 15740118, 16077202, 18540238). 
HK is supported by the 21st Century COE Program of Origin and
 Evolution of Planetary Systems in the Ministry of Education, Culture, 
 Sports, Science and Technology (MEXT) of Japan. 

%


\end{document}